# THE STATE OF THE ART IN HADRON BEAM COOLING

L.R. Prost[#], P. Derwent
FNAL[*], Batavia, IL 60510, USA

*Abstract*

Cooling of hadron beams (including heavy-ions) is a powerful technique by which accelerator facilities around the world achieve the necessary beam brightness for their physics research.

In this paper, we will give an overview of the latest developments in hadron beam cooling, for which high energy electron cooling at Fermilab's Recycler ring and bunched beam stochastic cooling at Brookhaven National Laboratory's RHIC facility represent two recent major accomplishments. Novel ideas in the field will also be introduced.

## INTRODUCTION

The provision for beam cooling capabilities is an important step in the conception and design of many accelerator facilities around the world. Depending on the application and the accelerator main parameters (energy, type of particles…), various techniques may be implemented. In addition, new cooling methods continue to be devised for future facilities.

While this paper will attempt to give an overview of the efforts carried out in the field as a whole, it will focus on the facilities which do use state-of-the-art cooling systems presently. In particular:

- The demonstration (and operation) of stochastic cooling for high-energy bunched beams at Brookhaven National Laboratory (BNL).
- The commissioning and cooling results from two state-of-the-art low energy coolers at the Institute of Modern Physics (IMP) Lanzhou in China and the European Organization for Nuclear Research (CERN) in Geneva, Switzerland.
- The operation and optimization of high-energy electron cooling (8 GeV antiprotons) at Fermi National Accelerator Laboratory (FNAL).

Also, while muon cooling is an effervescent topic (where information about recent activities can be found in Ref. [1-4] and within), it will not be covered in this paper which confines its scope to hadrons only.

## STOCHASTIC COOLING

### Coasting beams

Stochastic cooling techniques have been used successfully for decades and are now part of the standard tools for cooling hadron beams. The latest advances rely on the choice of the electronics and design of the pickups and kickers.

For instance, at FNAL, where stochastic cooling is used extensively, there are 12 cooling systems for 3 storage rings: 3 in the Debuncher (horizontal, vertical and longitudinal), 5 in the Accumulator (horizontal, vertical and 2 longitudinal – in different frequency bands - for core cooling plus the stacktail system) and 4 in the Recycler (2 horizontal – in different frequency bands-, vertical and longitudinal). In the Debuncher, an optical notch filter (with a depth of more than 30 dB) was recently installed and continues to be optimized. In the Accumulator where the stacking of antiprotons takes place, new equalizers were installed on 4 of the 5 systems (the last system will be upgraded soon) [5]. They are used to compensate the frequency response of the cooling systems, in particular the phase of the transfer function, which needs to be extremely flat. One feature of this new type of equalizer is that it separates the phase equalizer from the amplitude equalizer, each part being tunable independently. Finally, in the Recycler, the particularity of the cooling systems lies in the fact that the pickup signals (amplitude modulated) are carried across the ring to the kickers via a laser link light going through an evacuated pipe. Additionally, it is possible to gate two of the cooling systems such as to provide stochastic cooling over a fraction of the beam (isolated by RF barrier buckets).

### Bunch beams

Stochastic cooling of short bunches at high energies was attempted unsuccessfully some time ago in the Super Proton-Antiproton Synchrotron [6] and the Tevatron [7]. The main reasons for these failures were the very high coherent component of the Schottky signal that needs to be eliminated (without attenuating the Schottky 'noise') and the power requirements for the kicker(s). These two feats were recently accomplished at BNL in the Relativistic Heavy Ion Collider (RHIC) [8,9].

In order to attenuate the coherent signal, the Schottky spectrum is filtered before any amplification is applied. The filter consists of a series of coaxial cables, precisely timed to 5.000 ns intervals, splitters and combiners and is schematically depicted in Figure 1.

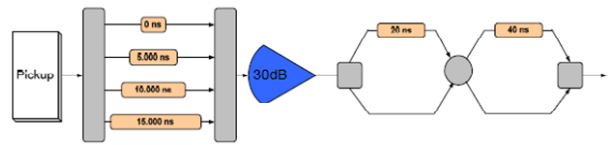

Figure 1: Traversal filter that reduces peak voltages from the pickup before the low noise amplifier and electrical-to-optical converter [9].

Then, the signal is amplified and sent to the 'kicker' through an externally modulated analog fiber optic link.

A total power of ~90 kW would be needed for a single kicker, with typical 50 Ohm load, to efficiently cool gold ions at 100 GeV/nucleon. In order to reduce the power needed, the kick is synthesized from 16 high-Q cavities spaced at 200 MHz intervals (bunch length is 5 ns) in the 5 to 8 GHz band of the system. The bandwidth of the cavities is chosen to allow filling and emptying the cavities between bunches (100 ns). Each cavity is driven by a 40 W solid state amplifier.

The longitudinal stochastic cooling system outlined above was installed in the yellow RHIC ring, tested with protons (at low intensity) and used operationally with gold ions [8-10]. As a result, the gold ions lifetime improved to the level close to the burn-off rate, while this is not the case for the blue RHIC ring, where stochastic cooling was not implemented (Figure 2).

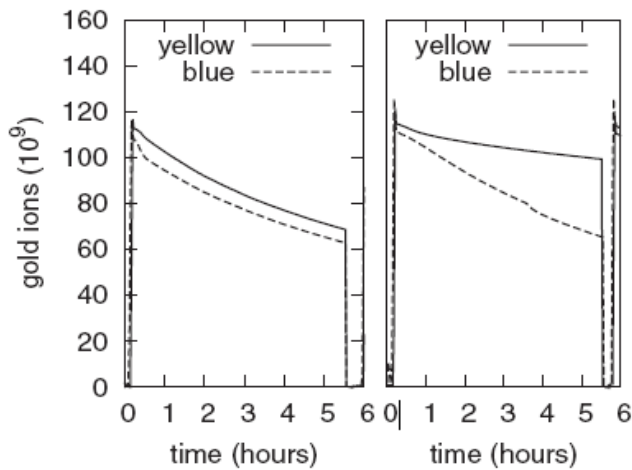

Figure 2: Evolution of the total beam intensity in RHIC for two physics stores in the yellow (solid line) and blue (dashed line) rings. The store on the left had no stochastic cooling while the one on the right had stochastic cooling on in the yellow ring [8].

Following the success encountered in one ring, the second ring (blue) was equipped with a similar system (longitudinal), which will be operational for the upcoming run (end of 2008). For the longer term, new transverse pickups are being designed for implementation of transverse cooling.

However, one problem remains to be addressed. While the bunch intensity remains quasi constant throughout a store length, under stochastic cooling the adjacent buckets get filled with beam. A couple of remedies have been identified: a quadrupole damper to damp the transition mismatch and a $2^{nd}$ harmonic cavity [11]. Both items will take some time to come to maturity.

*Optical stochastic cooling (OSC)*

One of the limitations of stochastic cooling systems using microwave signals is the bandwidth of RF amplifiers, in particular when it comes to intense bunched beams. To alleviate this limitation, the use of an optical system with much larger bandwidth was proposed [12]. The so-called transit-time method of OSC [13] uses two undulators through which the stored beam circulates: one to induce the optical radiation (like light sources), the second to apply momentum kicks back to the beam after amplification of the optical signal.

While this method was devised more than 10 years ago, no experimental proof-of-principle has been done. However, with the recent successes of X-FEL technologies, the OSC technique may become more practical and attractive. Thus, an experimental demonstration was proposed by the Massachusetts Institute of Technology-Bates (MIT-Bates) Linear Accelerator Center, using stored electrons [14]. A 3-year plan has been drawn up and is currently awaiting the appropriate funding to begin the construction of the necessary equipment.

*Lattice optimization*

While not a novel idea, a renewed interest has arisen in the design of 'optimum mixing rings' [15]. These 'asymmetric' lattices circumvent the 'mixing dilemma' by combining low dispersion from the pickup to the kicker (in the direction of the circulating beam) with high dispersion in the rest of the ring. One possible modular approach would be to construct cooling ring lattices consisting of small momentum compaction modules from pickup to kicker and negative momentum compaction modules from the kicker to the pickup.

Even in existing rings, it is sometimes possible to adjust the optics favorably for stochastic cooling. This was recently done at FNAL in the Accumulator, where the slip factor was increased by 15% in order to improve the stacktail cooling system efficiency (allowing for a higher antiproton flux) [16]. Although this is not a direct application of an 'asymmetric' lattice, the reasoning behind this optimization is the same.

## ELECTRON COOLING

While no new coolers were built over the past two years, there has been a lot of progress made at facilities which had just started operating state-of-the-art electron coolers, such as the ones at CERN for LEIR, at IMP Lanzhou for the CSRm ring (low energy) and at FNAL on the Recycler ring (high energy).

*High energy electron coolers*

As of now, FNAL's Recycler Electron Cooler (REC) [17,18] remains unique and its reliability over the past two years has been exceptional. Under normal conditions, its availability is close to 100%, with several day (2-3) maintenance/repair periods every 6 months or so. As a result, accumulation of $400\times10^{10}$ antiprotons (~3 times what would have been possible without electron cooling) is routinely achieved in support of the collider program.

The REC is based on a 4.3 MV electrostatic accelerator (Pelletron [19]) which works in the energy recovery mode (i.e. electrons are decelerated and captured in the collector after they interact with the antiproton beam). While recirculation of up to ~600 mA of DC beam current was achieved, for high voltage stability purposes cooling has been optimized at 100 mA, which has proved to be adequate thus far. Typical longitudinal 'normalized' cooling rates (recalculated to account for different momentum spreads of the antiproton beam) are shown on Figure 3.

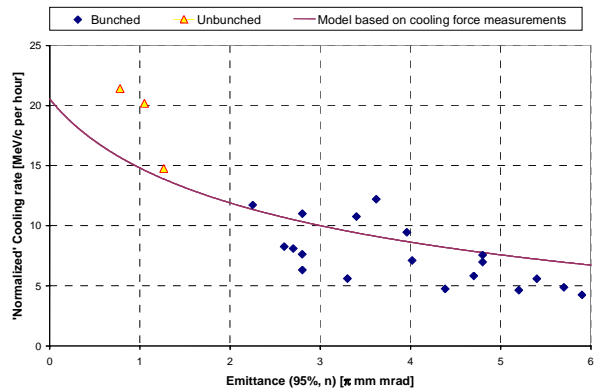

Figure 3: Longitudinal cooling rate as a function of the antiproton beam emittance.

More than the absolute value of the cooling rate, what is important on Figure 3 is its dependence on the antiproton beam emittance. Thus, in practice, stochastic cooling is continuously used in order to bring the transverse emittance of the beam to a level where electron cooling is the most efficient (longitudinally first but also transversely).

Other proposals at the Gesellschaft für Schwerionenforschung (GSI) for the High Energy Storage Ring (HESR) of the Facility for Antiproton and Ion Research (FAIR) and at BNL for the RHIC beams rely on similar designs as the one chosen for the Recycler ring (i.e. energy recovery scheme, long interaction regions). They differ in that a strong magnetic field is accompanying the electron beam from the cathode to the collector to enhance cooling through magnetization of the electron beam, as in low energy coolers (for which the strong magnetic field is required to counteract the space-charge of the beam). This is not the case at FNAL, where electron cooling is described by the 'non-magnetized' friction force model. The latest versions of the BNL design dropped the requirement for magnetization of the electron beam after FNAL's demonstrated success with un-magnetized cooling.

However, the success of stochastic cooling in RHIC led to the cancellation of the electron cooling project for high energies at BNL. On the other hand, discussions continue about the possibility of using electron cooling at lower energies ($\gamma \sim 2.6$-$12$) for which a quasi-replica of the REC would be adequate. Meanwhile, FAIR's high energy electron cooling project is moving forward with a recent test of a single HV section designed and built by the Budker Institute for Nuclear Physics (BINP) in Novosibirsk, Russia [20]. At this stage, it is foreseen that the 2-MeV cooler proposal for the Cooler Synchrotron (COSY) [21] at Forschungszentrum Jülich, Germany, would be an intermediate step towards the final design and construction of the cooler for the HESR ring.

*Low energy coolers*

As reported elsewhere [22-24], the latest generation of low energy coolers (up to ~300 keV electron energy) at both IMP and CERN feature several innovations intended at improving upon difficulties encountered with the first generation coolers. These are (in no particular order): a very precise magnetic field in the cooling section obtained through the use of a large number of trim coils and high perveance in order to attain faster cooling; electrostatic bends to reduce trapping of secondary particles; magnetic expansion to adjust the beam size; beam shaping capabilities ('hollow beam') to help reduce 'overcooling' of the core particles and reduce ion-electron recombination. While some features such as the magnetic field quality and the push to higher beam current are known to improve cooling capabilities, others, like beam shaping, are more controversial as to their benefits.

Along with the standard commissioning activities (which have not been entirely completed to this date), specific measurements were carried out to investigate the efficiency of some of these new features on the cooling process and the electron beam stability. For instance, CERN reported on the positive influence of the electrostatic bend on the maximum beam current that can be extracted [25] and the influence of the electron beam size on cooling efficiency, which, they find, peaks when the electron beam has roughly the same size as the ion beam (Figure 4) [26].

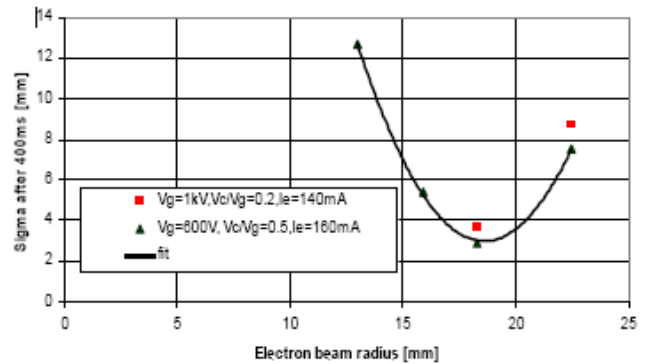

Figure 4: Beam size 400 ms after first injection as a function of the electron beam radius [26].

Both IMP and CERN also investigated the effect of cooling with a hollow beam, and this will be discussed in a later section.

The purpose of the cooler in both cases was the overall improvement of accumulation: in preparation for injection of lead ions into the Large Hadron Collider (LHC) at LEIR, for physics experiments in the CSRe ring at the CSRm ring. Through lifetime improvements with electron cooling, LEIR was able to deliver the required beam parameters for the first LHC ion run ($N_{Pb} = 2.2 \times 10^8$, $\varepsilon_{h,v} < 0.7$ μm in a 3.6 s cycle) while the CSRm saw its accumulation capability increase by one order of magnitude with smaller transverse and longitudinal emittance (by ~10 w.r.t. without electron cooling).

*Stability of cooled beams*

It is now well known and documented (a summary can be found in Ref. [27]) that a well cooled ion beam may manifest an increased sensitivity to instabilities of various origins (beam-beam, 'three-body' instability when secondary ions get trapped in the electron beam potential, coupling impedance).

What may be the more problematic are the 'beam-beam' effects that appear to be leading to slow ion loss (or at least some additional diffusion) or fast losses at injection (under certain conditions). Although some theoretical considerations have been brought forward [24], a deep understanding of the mechanisms involved is still lacking. Nonetheless, the experimental data suggest that the losses are due to 'overcooling' of the center of the ion beam. In fact, at FNAL, the cooling rate is adjusted by shifting the electron beam in and out of the antiproton beam rather than by changing the electron beam current. One of the reasons is that when the electron beam (which diameter is usually larger than the rms size of the antiproton transverse distribution) completely overlaps the antiproton beam, the antiproton lifetime deteriorates very quickly. During accumulation, adequate cooling is provided with good lifetime with the electron beam offset by a couple of millimeters with respect to the closed orbit of the antiprotons. On the other hand, just before extraction (e.g.: minutes), when the antiproton lifetime is no longer a primary concern, the electron beam is positioned such as to maximize cooling (i.e. brought collinear to the antiproton beam closed orbit).

Similarly, the design of low energy coolers with beam shaping capabilities arose from the 'overcooling' of the core problem. By allowing the electron beam density distribution to vary across its cross-section (from 'low' near the center to 'high' at the edge), it was thought that the density of the cooled beam would be more uniform, hence avoiding (or reducing) its susceptibility to losing particles. Practically, the design of the gun includes a control electrode which shapes the electric field near the cathode. The ratio of the control electrode voltage ($U_{ce}$) to the anode voltage ($U_a$) is what determines the current density profile.

Preliminary measurements at LEIR [25,26,28] are somewhat inconclusive at this point. While the use of a slightly hollow beam allowed nearly doubling the number of lead ions to be stacked for extraction to the Proton Synchrotron (PS) [25], some dedicated cooling studies [26] indicated that the electron beam density distribution had only a very small effect (if any) on the cooling efficiency or the long-term (~10 s) lifetime. It was suggested [29] that a possible reason for this apparent discrepancy may be the effect that cooling with a hollow beam has on the short-term lifetime (1-2 s), which is what matters in LEIR's 3.6-s injection cycle.

More systematic measurements carried out on the CSRm ring showed better accumulation results on a number of ions when the ratio of the control electrode voltage to the anode voltage is ~0.2-0.4 [28,30]. This is illustrated in Figure 5 where an image of the electron beam current density distribution with $U_{ce}/U_a \approx 0.2$ is shown along with measurements of maximum accumulation versus 'hollowness' of the electron beam.

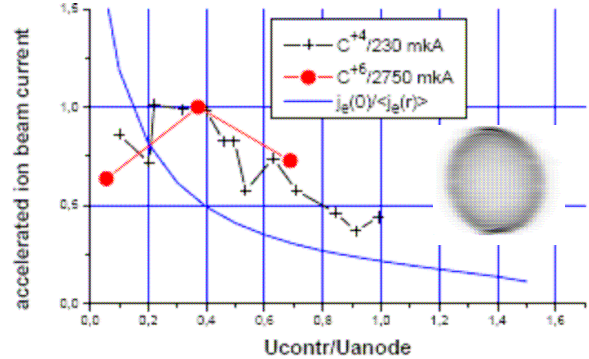

Figure 5: Maximum ion beam currents accelerated at CSRm (for $C^{4+}$ and $C^{6+}$ ions) vs. the ratio of control to anode voltage. Blue line: Ratio of the electron beam density at the center to the average density (for a flat beam j(0)/<j(r)>=1) [28].

In both cases (LEIR and CSRm), more measurements will be carried out in the near future in order to address in a more definite way the benefits (or lack thereof) of cooling ion beams with a hollow electron beam.

*Electron cooling of very low energy ions*

ELENA [31] at CERN and FLAIR [32] at FAIR are two proposals that require efficient cooling of ~100 keV antiprotons and, in the case of FLAIR, slow ions too. In parallel, electrostatic cooling rings have been constructed or are under construction (e.g.: the Cryogenic Storage Ring (CSR) [33] at the Max-Planck Institut für Kernphysik (MPI-K) in Heidelberg, Germany or DESIREE [34] at the Manne Siegbahn Laboratory in Stockholm, Sweden).

One of the challenges for electron cooling is the ultra low temperature (a few meV) of the electron beam which is required. In order to achieve such a temperature, the use of a 'cold' photocathode has been proposed and was tested at MPI-K on the Test Storage Ring (TSR) [35]. Using a 53 eV electron beam produced by a cryogenic GaAs photocathode, they successfully and rapidly (~2s transverse cooling time) cooled down 97 KeV/nucleon $CF^+$ ions to very low emittances, showing the potential of cryogenic photocathodes for electron cooling of slow ions.

## BEAM ORDERING EXPERIMENTS

Beam ordering experiments refers to experiments aimed at achieving conditions such that ions in the beam re-arrange themselves in a crystal-like structure. It was demonstrated in GSI for 1D ordering [36] and later in the PALLAS ring at the University of Munich, Germany, 3D ordering was achieved at very low energy [37].

At the Institute of Chemical Research (ICR), Kyoto, Japan, the S-LSR ring was constructed and commissioned [38] with the purpose of achieving 3D ordering. Its main feature is the possibility to run in a dispersion free mode thanks to specially designed bending fields that are obtained with both magnetic and electrostatic fields. The purpose of designing a dispersion free lattice was to eliminate 'shearing forces' identified as one major obstacle for 2D and 3D ordering at high energies. In addition, the ring is equipped with both an electron cooler and a laser cooling system.

The first set of experiments at the ICR focused on electron cooling of protons and 1D ordering of 7 MeV protons was achieved [39]. This is illustrated on Figure 6 where the momentum spread of the beam and the total Schottky noise drop abruptly under strong cooling once the number of particles is less than ~2000.

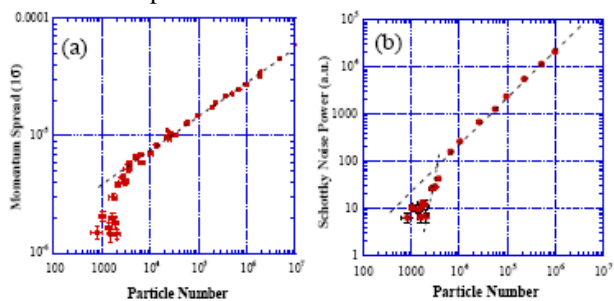

Figure 6: Momentum spread (a) and Schottky noise power (b) as a function of the particle numbers with the electron current of 25 mA [39].

More recently, the focus of the experiments shifted to the laser cooling technique for which the S-LSR is optimized to realize multi-dimensional crystalline beams. Magnesium ions were used in laser cooling experiments for which two methods were tried: in the first method the ion beam is decelerated by an induction voltage while keeping the laser frequency constant; in the second the laser frequency was swept across the velocity spread of the ion beam population [40,41]. 1D ordering has not been observed yet and steps towards this goal are being implemented (in particular, cooling in the transverse direction since laser cooling is longitudinal only). In addition, note that none of the experiments carried out so far were performed using the dispersion free configuration (which is not required for 1D ordering).

## NUMERICAL SIMULATIONS

In order to design and build new cooling systems, adequate simulations of the beam dynamics must exist. For the cooling processes (and related effects such as intra-beam scattering, gas scattering…), the BETACOOL code [42] has matured and is now quite widely used. Meanwhile, benchmarking and improvements/additions to the code are continuously made.

There are basically two sorts of calculations that can be performed: 'rms dynamics' or 'Model Beam' [43]. In the 'rms dynamics' algorithm, the evolution of the second order momenta of the ion distribution function is calculated. This calculation is based on the assumption of Gaussian distributions and all heating and cooling effects are characterized by rates of variation of the emittances or of particle loss. For an arbitrary ion distribution, a multi-particle simulation algorithm is employed (i.e. 'Model Beam' algorithm). In this case, the ion beam is represented by an array of model particles for which the heating and cooling processes lead to changes of the components of the particles momentum and of the number of particle.

Some of the recent additions to the code (not all completed) are stochastic cooling processes (including bunched beams for RHIC), longitudinal dynamics with barrier buckets, detailed IBS models and laser cooling.

In parallel to the development of the BETACOOL code, 'direct' calculation of binary collisions of ions with cooling electrons are performed using the framework of the VORPAL code [44]. The goal is to improve the models used in BETACOOL, for instance, in the computation of the friction force [45,46].

## R&D PROJECTS ON FUTURE FACILITIES

There are currently three or four major projects for which extended and innovative cooling systems will be needed.

At GSI first, the FAIR project includes several storage rings, each of these needing some sort of cooling system. At this stage, stochastic cooling appears to be the front-runner, with the necessary addition of an electron cooler in the HESR. So far, efforts have been put in understanding what are the requirements for stochastic cooling and their implication in the choice and design of the subsystems (pick-ups, amplifiers, kickers). Various pickup designs have been tested for the HESR [47] and the Cooling Ring (CR) [48], as well as power amplifiers. As for the electron cooler, as noted earlier, the development of the accelerator currently relies on the activities carried out at BINP in collaboration with FJZ, while the University of Uppsala, Sweden, focuses on the electron gun and collector systems.

At BNL, two projects are under investigation and would need electron cooling. The first is 'critRHIC' [49,50], a low-energy RHIC operation ($\gamma \sim 2.6\text{-}12$) for which a cooler similar to FNAL's would probably be sufficient. The second project is an electron-ion collider 'eRHIC' [51] for which a totally new electron cooler design would be necessary, based on an Energy Recovery Linac (ERL) for the generation of the electron beam. At the same time, BNL is now considering using Coherent Electron Cooling (CEC) [52,53], which in theory could improve cooling rates by several orders of magnitude for high-energy, high-intensity ion beams. A proof-of-principle experiment at RHIC might be proposed for the middle of the next decade, following the commissioning of their ERL currently under construction [54].

An electron-ion collider proposal [55] is also under consideration at Jefferson Laboratory (JLab), for which a conceptual design of the electron cooler has been completed [56]. It would be made of two parts: a 30 mA, 125 MeV ERL operating at 15 MHz and feeding a 3 A circulator-cooler ring (CCR) operating at 1500 MHz bunch repetition rate (Figure 7).

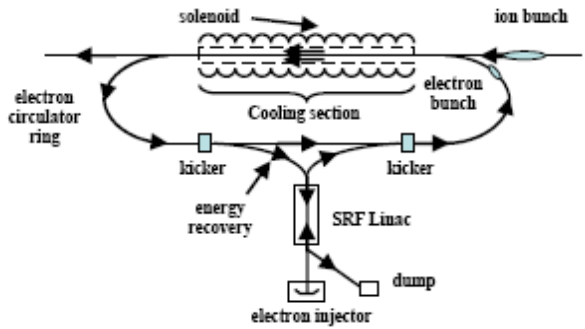

Figure 7: Layout of the electron cooler for ELIC [56].

The collider itself would be composed of two figure-eight rings with 4 interaction regions [56,57].

## CONCLUSION

The field of beam cooling is wide and continues to flourish with new facilities being designed and old ideas being pursued. The work, accomplishments and plans presented in this report by no way constitute a complete summary of everything that is being done in the field today. It merely highlights what the authors, inevitably biased, judged of prime interest.

More detailed information can be found in the Proceedings of the Workshops on Beam Cooling and Related Topics held every two years, from which many of the references in this paper were extracted from.

## ACKNOWLEDGMENTS

We would particularly like to thank V. Parkhomchuk, G. Tranquille, J. Dietrich, A. Fedotov and M. Blaskiewicz, for their help on gathering some of the relevant information for the preparation of this report and sharing their expertise in the field.